\documentclass[sigconf,nonacm]{acmart}
\AtBeginDocument{%
  }

\setcopyright{none}
\copyrightyear{}
\acmYear{}
\acmDOI{}
\acmConference[~]{~}{~}{~}
\acmBooktitle{~}
\acmISBN{~}

\usepackage{booktabs}
\usepackage{multirow}
\usepackage[table,xcdraw]{xcolor}
\usepackage{enumitem}
\usepackage{balance}

\usepackage{makecell}
\usepackage{minted}
\setminted{
    fontsize=\footnotesize,
    linenos=true,
    breaklines=true,
    breaksymbolleft={},
    xleftmargin=2em,
}

\usepackage{listings}
\usepackage{xcolor}

\lstdefinelanguage{MLIR}{
    keywords={
        module, func.func, return
    },
    morekeywords={
        affine.for, affine.load, affine.store,
        arith.muli
    },
    sensitive=true,
    morecomment=[l]{//},
}

\lstdefinelanguage{SystemVerilogCustom}{
    language=Verilog,
    morekeywords={
        logic, genvar, generate, endgenerate,
        signed
    },
    sensitive=true,
    morecomment=[l]{//}
}

\usepackage[most]{tcolorbox} 
\usepackage{xcolor}

\usepackage{xcolor}

\definecolor{bg_gray}{RGB}{245,245,245}
\definecolor{bg_blue}{RGB}{240,248,255}
\definecolor{bg_green}{RGB}{240,255,240}
\definecolor{frame_gray}{RGB}{180,180,180}

\newtcblisting{icmlcode}[2][]{
    enhanced,
    breakable,
    colback=bg_gray,
    colframe=frame_gray,
    boxrule=0.5pt,
    arc=2mm,
    fonttitle=\bfseries\sffamily,
    title={#2},
    #1, 
    listing only,
    listing options={
        language=Bash,
        basicstyle=\ttfamily\footnotesize,
        keywordstyle=\color{blue!70!black}\bfseries,
        commentstyle=\color{green!40!black},
        stringstyle=\color{orange!90!black},
        numbers=left,
        numberstyle=\tiny\color{gray},
        stepnumber=1,
        breaklines=true,
        columns=fullflexible,
        tabsize=4,
        showstringspaces=false
    }
}

\newtcblisting{icmlprompt}[2][]{
    enhanced,
    breakable,
    colback=bg_blue,
    colframe=frame_gray,
    boxrule=0.5pt,
    arc=2mm,
    fonttitle=\bfseries\sffamily,
    title={#2},
    #1,
    listing only,
    listing options={
        basicstyle=\ttfamily\footnotesize,
        breaklines=true,
        columns=fullflexible,
        escapechar=|  
    }
}

\begin{document}

\title{EquivFusion: Unifying Hardware Equivalence Checking from Algorithms to Netlists via MLIR}








\author{
Jiaying Zhu$^{1,3}$,
Baoqi Zhang$^{1}$,
Mengxia Tao$^{1}$,
Kezhi Li$^{3}$,
Hao Yan$^{1,2}$,
Qiang Xu$^{3}$,
Min Li$^{1,2}$
}

\authornote{Correspondence: min.li@seu.edu.cn}

\affiliation{
\institution{$^{1}$National Center of Technology Innovation for EDA, Nanjing, China}
\city{}\country{}
}

\affiliation{
\institution{$^{2}$Southeast University, Nanjing, China}
\city{}\country{}
}

\affiliation{
\institution{$^{3}$The Chinese University of Hong Kong, Hong Kong, China}
\city{}\country{}
}

\renewcommand{\shortauthors}{Jiaying Zhu et al.}



\begin{abstract}
  Ensuring functional consistency between high-level algorithmic models and low-level hardware implementations is a critical challenge, particularly as modern design flows increasingly span heterogeneous abstractions---from deep learning frameworks to hardware netlists. In this paper, we present \textsc{EquivFusion}, an end-to-end equivalence checking tool tailored for multi-modal circuit designs. Unlike traditional flows that rely on siloed tools or ad-hoc translation, EquivFusion leverages a verification-oriented MLIR lowering pipeline to unify diverse entry points---including PyTorch, C/C++, Chisel, Verilog, and gate-level netlists---into a common intermediate representation. This architecture enables automated, pairwise equivalence checking across diverse abstraction levels by rigorously translating designs into standard formal verification formats, i.e., SMT-LIB, BTOR2, AIGER. We demonstrate EquivFusion's feasibility to bridge the semantic gap between software specifications and hardware realizations, showcasing its effectiveness in facilitating “shift-left” formal verification for datapath-intensive hardware designs. 
  EquivFusion is available online\footnote{Code: \url{https://github.com/FORMiND-Lab/EquivFusion}}.

\end{abstract}

\maketitle

\section{Introduction}
\label{sec:intro}

The complexity of modern hardware systems has necessitated a paradigm shift in design flows. To cope with the demands of domain-specific accelerators and data-intensive workloads, designers are moving beyond traditional Register-Transfer Level (RTL) entry points. The contemporary design spectrum is highly heterogeneous: algorithmic specifications are often defined in deep learning frameworks like PyTorch~\cite{imambi2021pytorch}; high-level synthesis (HLS) models~\cite{coussy2010high} are written in C++; and hardware construction languages (HCLs) such as Chisel~\cite{bachrach2012chisel} are used to generate RTL, which is subsequently synthesized into gate-level netlists for implementation.
This diversity enables rapid innovation but introduces a fundamental challenge: ensuring functional correctness across representations that differ drastically in semantics, granularity, and tooling support.

For example, verifying that a PyTorch model faithfully corresponds to its Verilog counterpart is critical in domains like mobile System-on-Chips (SoCs), where camera pipelines increasingly replace traditional Image Signal Processors (ISP) with end-to-end deep neural hardware~\cite{schwartz2018deepisp}. Similarly, datapath-intensive accelerators for vision or language models often originate from high-level algorithmic intent but must be validated against low-level realizations~\cite{coussy2010high, chen2018tvm}. These scenarios highlight the need for robust equivalence checking across heterogeneous design abstractions~\cite{wang2023equivalence}.

Existing equivalence checking tools nevertheless remain largely confined to specific abstraction boundaries and design modalities. Open-source engines such as ABC~\cite{brayton2010abc} and Yosys EQY~\cite{Wolf2013YosysAFV, yosys_eqy} primarily support logic equivalence checking for RTL and gate-level designs, while HW-CBMC~\cite{hwcbmc2017} targets bounded checking between C and Verilog. Commercial tools such as Synopsys VC Formal Datapath~\cite{synopsys_vcformal_dpv} and Cadence C2RTL~\cite{cadence_jasperc_formal} focus on C/C++-to-RTL datapath validation, and Synopsys Formality~\cite{synopsys_formality} is widely used for RTL-to-gate equivalence in synthesis sign-off. Although effective within their intended scopes, these tools do not provide an extensible end-to-end front-end that can directly ingest modern algorithmic specifications, such as PyTorch models, together with hardware implementations, such as Verilog or netlists, and align them for equivalence checking. Consequently, cross-modal validation often still relies on simulation-driven flows~\cite{Simulation2011, RTLTLM2008}, which require manually developed testbenches and can miss subtle arithmetic mismatches in datapath-intensive designs~\cite{blum2002reflections, drechsler2022polynomial}.

The emergence of MLIR~\cite{MILR2021, mlirllvm} and CIRCT~\cite{circt} offers a promising foundation for unification. Their multi-level design allows structural and semantic information to be captured explicitly across abstraction boundaries. However, existing MLIR workflows are primarily oriented toward compilation, lacking a verification-oriented pipeline capable of modeling equivalence relations or generating proof obligations. While recent work~\cite{dobis2024formal} has explored embedding logic equivalence checking (LEC) and a small subset of SystemVerilog Assertions (SVAs)~\cite{das2006synthesis} into MLIR, it  does not address the broader challenge of cross-modal hardware equivalence checking. As verification shifts earlier in the design cycle, there is a critical need for a system that leverages MLIR's representational power while automating the generation of formal checks.

To this end, we introduce \textsc{EquivFusion}, an equivalence verification pipeline that unifies heterogeneous design modalities. We leverage MLIR not merely as a compilation infrastructure, but as the backbone of a verification workflow. EquivFusion advances the state of hardware verification through the following contributions:

\begin{itemize}[itemsep=0.2em, leftmargin=*]
    \item \textbf{Unified Multi-Modal Frontend:} \textsc{EquivFusion} acts as a semantic bridge, enabling automated equivalence checking between diverse abstraction levels. Users can formally verify a Chisel module against its C++ behavioral specification, or a synthesizable PyTorch model directly against its RTL implementation.
    \item \textbf{Verification-Oriented Lowering:} We implement a specialized MLIR pipeline that preserves high-level semantics (e.g., array operations, loop structures) where beneficial, while progressively lowering hardware specifics to a canonical representation. This avoids the semantic loss often associated with generic synthesis.
    \item \textbf{Flexible Multi-Engine Backend:} Rather than binding to a single engine, EquivFusion exports designs to standard formats including SMT-LIB~\cite{barrett2010smt}, BTOR2~\cite{niemetz2018btor2}, and AIGER~\cite{biere2007aiger}. This flexibility allows users to leverage best-in-class open-source solvers (e.g., Z3~\cite{de2008z3}, Bitwuzla~\cite{niemetz2023bitwuzla}, kissat~
    \cite{fleury2020cadical}) tailored to the specific logic depth of the problem.
\end{itemize}
Ultimately, \textsc{EquivFusion} bridges the semantic gap between specifications and implementations, enabling automated “shift-left” verification that traps functional errors early in the design cycle.
\section{Related Work}
\label{sec:related_work}

\subsection{MLIR-based Hardware Infrastructures}
MLIR~\cite{MILR2021, mlirllvm} serves as a reusable compiler infrastructure designed to support domain-specific intermediate representations (IRs) through a unified dialect system. Its key innovation lies in enabling IRs at multiple abstraction levels to coexist and interact, facilitating progressive lowering and improving interoperability with downstream EDA tools. Built on MLIR, CIRCT~\cite{circt} (Circuit IR Compilers and Tools) provides a collection of hardware-oriented dialects and tooling intended to form modular, reusable flows and to interoperate with downstream EDA tools. Recent work~\cite{dobis2024formal} implements basic logic checks and assertions into CIRCT, yet it is limited to small-scale reasoning and does not address the semantic gap between high-level algorithmic models and hardware implementations.


\subsection{Hardware Equivalence Checking}
Equivalence checking is a central technique in hardware verification for ensuring functional consistency across design transformations~\cite{marques2000boolean, Kuehlmann2002}. 
Compared to standard flows where designs stabilize within a single domain, checking equivalence across abstraction levels-particularly between software reference models and hardware implementations-remains more challenging. 
It is commonly addressed through simulation-based verification or manually maintained reference models, approaches incurring significant engineering effort and scale poorly for datapath-intensive designs.

Existing tools focus on limited verification scenarios. Commercial solutions such as Synopsys Formality~\cite{synopsys_formality} and Cadence Conformal~\cite{cadence_conformal} target RTL-to-netlist equivalence, while Synopsys VC Formal DPV~\cite{synopsys_vcformal_dpv} and Cadence C2RTL~\cite{cadence_jasperc_formal} support C-to-RTL validation.
Open-source tools including ABC~\cite{brayton2010abc}, the Yosys EQY flow~\cite{Wolf2013YosysAFV, yosys_eqy}, and HW-CBMC~\cite{hwcbmc2017} provide scalable boolean or bounded reasoning, but are tightly coupled to fixed language pairs and RTL/netlist flows, lacking extensible front-end support for modern algorithmic specifications and hardware languages such as PyTorch or Chisel.

\section{Overview of the Tool}

\subsection{Overview of Architecture}

\textsc{EquivFusion} addresses cross-modal equivalence checking between algorithmic specifications and hardware implementations by unifying heterogeneous inputs into a common intermediate representation.
To support “shift-left” verification, it adopts a verification-oriented lowering strategy that preserves reasoning-relevant semantics and delays bit-level refinement when possible.
The tool is designed for datapath-intensive modules dominated by arithmetic and structured dataflow (e.g., kernels and accelerator datapaths), while control-dominant protocols with rich temporal behaviors are out of scope and are better served by hardware model checking~\cite{griggio2015comparing}.

\begin{figure}[t!]
    \centering
    \includegraphics[width=0.9\linewidth]{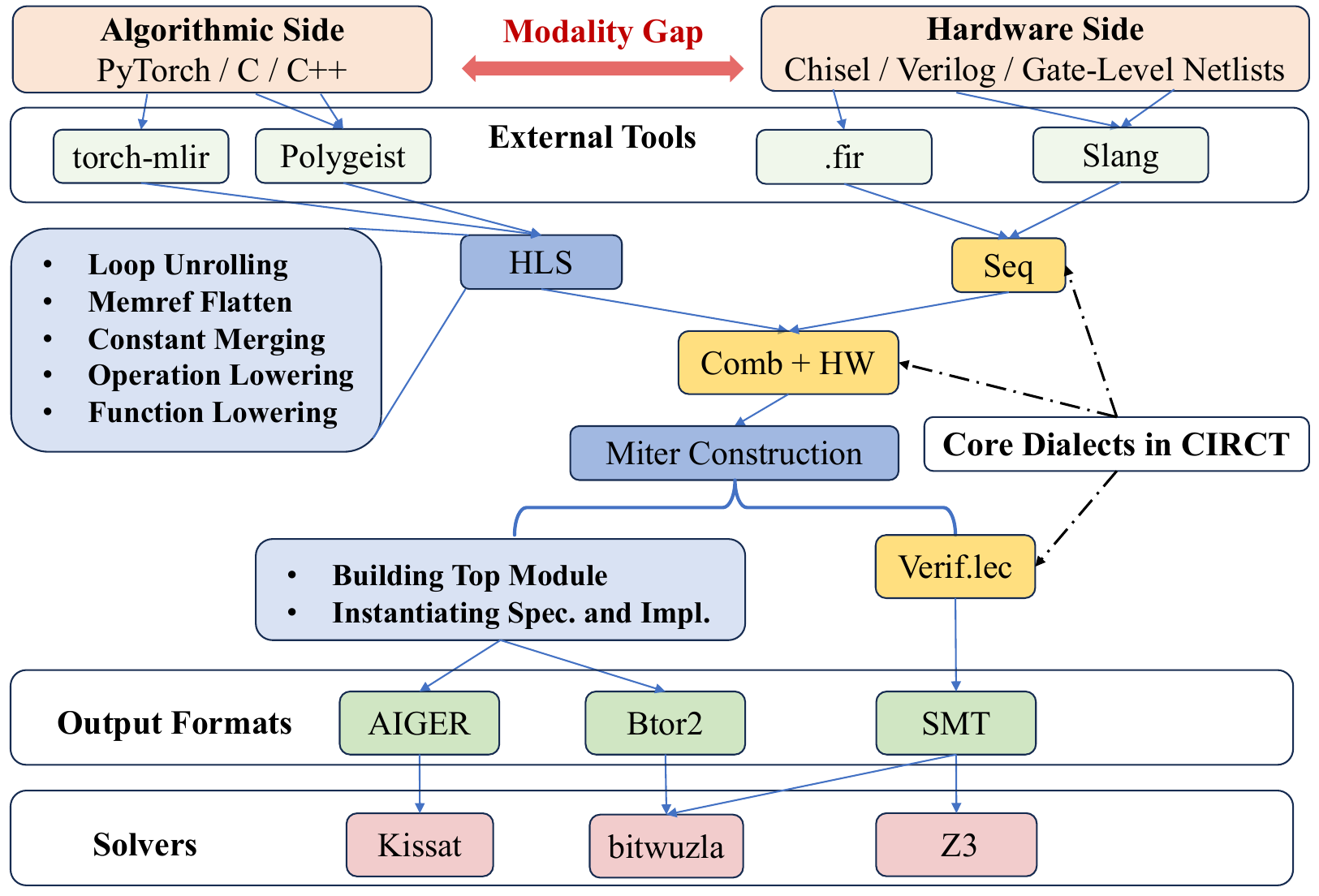}
    \vspace{-5pt}
    \caption{Overview of EquivFusion.}
    \label{fig:oveview}
    \vspace{-10pt}
\end{figure}

As shown in Figure~\ref{fig:oveview}, \textsc{EquivFusion} accepts designs from diverse entry points, including PyTorch, C/C++, Chisel, Verilog, and gate-level netlists.
Each input is first translated by existing front-ends into MLIR, bringing heterogeneous specifications into a common intermediate form.
At the core, \textsc{EquivFusion} hosts a set of verification-oriented dialects on top of CIRCT:
sequential designs are represented using a dedicated sequential abstraction and can be selectively unrolled to derive equivalent combinational views, while combinational and hardware-level structure is maintained through progressive lowering.
On the unified IR, \textsc{EquivFusion} constructs a miter circuit by instantiating the specification and implementation under shared inputs and encoding output equivalence constraints.
Finally, the miter is exported to standard solver formats (SMT-LIB, BTOR2, AIGER) and discharged to off-the-shelf SAT/SMT solvers.
This modular design decouples language front-ends from verification back-ends, making it straightforward to extend \textsc{EquivFusion} with new input modalities or solvers.

\subsection{Verification Scope}

\noindent\textit{3.2.1  Combinational Equivalence Checking.} 
\textsc{EquivFusion} supports combinational equivalence checking for designs that can be modeled as pure functions from inputs to outputs within a single cycle.
This setting covers a wide range of modules extracted from larger systems, including arithmetic kernels, feed-forward blocks, and bit-precise datapath components, which may arise as standalone designs or as combinational regions from larger sequential systems.

\vspace{2.5pt}
\noindent\textit{3.2.2  Transactional Equivalence Checking.} 
Beyond purely combinational settings, \textsc{EquivFusion} also targets equivalence checking for multi-cycle designs, where hardware behavior spans multiple clock cycles.
In such scenarios, the two implementations may differ in cycle-level timing, latency, or internal pipelining, yet are expected to produce equivalent results over a complete transaction from inputs to outputs.
Typical examples include validating an algorithmic reference model against a pipelined RTL implementation, or checking a high-level description with implicit scheduling against a multi-cycle hardware realization.

\section{Implementation Details}

\subsection{User Interface and Workflow}

The \textsc{EquivFusion} framework is accessed via its primary command-line interface (CLI), the \texttt{equiv\_fusion} utility. A typical hardware equivalence verification workflow consists of five sequential steps, each invoked by a specific subcommand:

\vspace{6pt}
\begin{lstlisting}[language=bash,  numbers=none]
# Step 1: Specify input/output ports for algorithmic design
set_port <-input|-output> port-name
# Step 2: Read specification design
read_c|read_v|read_firrtl -spec -top <module-name> <input-file>
# Step 3: Read implementation design
read_c|read_v|read_firrtl -impl -top <module-name> <input-file>
# Step 4: Construct miter and generate output file
equiv_miter -specModule <module-name> -implModule <module-name> -mitermode <smtlib|btor2|aiger> -o <miter-file>
# Step 5: Run solver for verification
solver_runner -solver <solver-name> -inputfile <miter-file>
\end{lstlisting}
\vspace{-10pt}
\captionof{listing}{Execute Steps}
\label{code:execute_steps}

\subsection{Unified Multi-Modal Frontend}
To bridge the modality gap between heterogeneous abstraction levels, \textsc{EquivFusion} implements specialized translation flows built upon the CIRCT infrastructure. 
These flows map both algorithmic (Pytorch/C/C++) and hardware (Chisel/Verilog/netlists) descriptions onto a unified intermediate representation comprised of the CIRCT core dialects (HW, Comb, Seq, and Verif).

\vspace{2pt}
\noindent\textit{4.2.1  Algorithmic Description Translation.} 
For high-level languages like C/C++ and PyTorch, \textsc{EquivFusion} targets a specific synthesizable subset conducive to High-Level Synthesis (HLS). 
\textsc{EquivFusion} leverages external compiler infrastructures to convert high-level specifications into the MLIR Affine dialect. Specifically, Polygeist~\cite{polygeistPACT} translates C/C++ programs, while torch-mlir~\cite{torch-mlir} and upstream MLIR tooling are used to convert PyTorch models.

With the Affine dialect as the entry point, \textsc{EquivFusion} implements a custom HLS flow built upon the CIRCT infrastructure. This flow progressively lowers the Affine IR by applying a sequence of both existing CIRCT transformations and custom-implemented passes. The flow executes key operations including Loop Unrolling, Memref Flattening, Constant Merging, and Operation/Function Lowering, ultimately converting the high-level representation into an IR expressed in CIRCT's core dialects.

\vspace{2.5pt}
\noindent\textit{4.2.2  Hardware Description Translation.} 
For Chisel designs, \textsc{EquivFusion} utilizes firtool (a utility within CIRCT) to parse the \textit{.fir} file generated by the Chisel compiler and convert it into an IR expressed in CIRCT's core dialects. For Verilog/Netlist inputs, the tool employs \textit{circt-verilog} (also within CIRCT), which leverages the Slang parser to process the source designs  broad syntactic support. Both paths ultimately convert the hardware descriptions into the unified IR expressed in CIRCT's core dialects.

Notably, \textsc{EquivFusion} implements sequential unrolling to handle state-dependent logic. 
This technique expands the sequential circuitry over a specified number of clock cycles ($k$) to construct a cumulative combinational model. 
This model is behaviorally equivalent to the original design over the $k$ unrolled time-frames, thereby enabling transactional equivalence checking to rigorously verify state transitions across a finite temporal window.

\subsection{Miter Construction Middle-End}
The \texttt{equiv\_miter} command orchestrates the miter circuit construction process, synthesizing a unified verification model from the distinct specification and implementation modules. The resulting miter circuit is subsequently exported into standard formats to facilitate formal equivalence checking.

\vspace{2pt}
\noindent\textit{4.3.1  Input/Output Mapping.} 
The fundamental premise of the equivalence verification model in \textsc{EquivFusion} is that, given identical input stimuli, both the specification and implementation modules must produce identical outputs. Consequently, prior to constructing the miter model, \textsc{EquivFusion} establishes a one-to-one I/O correspondence between the \textit{spec. module} and the \textit{impl. module} via a port matching mechanism. \textsc{EquivFusion} establishes a one-to-one correspondence between specification and implementation ports by enforcing consistency in port count, names, and types.
Upon successful validation, \textsc{EquivFusion} binds the ports by name to finalize the mapping for verification. Conversely, if any check fails, the tool reports a mismatch error and terminates the process.

\vspace{2.5pt}
\noindent\textit{4.3.2  Construct Miter.} 
\label{sec::construct_miter}
\textsc{EquivFusion} supports multiple modes for miter construction, including smtlib, btor2, and aiger.
Depending on the selected mode, equivalence constraints are encoded either as word-level assertions or bit-level XOR checks, enabling compatibility across SAT and SMT solvers.

\subsection{Solving Backends and Verification Results}
The \texttt{solver\_runner} command serves as the unified interface between \textsc{EquivFusion}'s internal IR and external formal verification engines. It automates the invocation of backend solvers tailored to the specific verification format generated in the previous stage.

\vspace{2pt}
\noindent\textit{4.4.1 Supported Solvers. } \textsc{EquivFusion} currently integrates three industry-standard solvers to cover different verification needs: Z3~\cite{de2008z3} (for SMT-LIB), Bitwuzla~\cite{niemetz2023bitwuzla} (optimized for BTOR2 word-level reasoning), and Kissat~\cite{fleury2020cadical} (for AIGER bit-level verification). This flexibility allows users to select the most efficient engine based on the circuit's complexity and abstraction level.

\vspace{2.5pt}    
\noindent\textit{4.4.2 Verification Outcome and Debugging. }  The solver's output provides a definitive conclusion regarding the functional consistency of the design pair: \textit{UNSAT (Equivalent)} indicates that no input combination exists that causes the outputs to diverge, proving that the implementation is functionally equivalent to the specification; \textit{SAT (Bugs Found)} indicates that the designs are not equivalent. Crucially, \textsc{EquivFusion} captures the \textit{counterexample} provided by the solver --- a specific assignment of input values that triggers the mismatch. This counterexample serves as a precise diagnostic trace, enabling designers to reproduce the error and pinpoint the logic bug within the hardware implementation.

\section{Usage Scenarios and Case Studies}



This section demonstrates the practical usage of \textsc{EquivFusion} through representative case studies.

\subsection{Example 1: C++ \textit{v.s.} Chisel} \label{sec:cpp_chisel}

\vspace{2pt}
\noindent\textit{5.1.1  Experimental Setup.} 
We verify a sorting module for unsigned integers to demonstrate equivalence checking across different abstraction levels. The setup contrasts a sequential software reference against a parallel hardware implementation:

\textbf{Specification (C++): } Bubble Sort algorithm (Listing~\ref{code:sort_cpp})

\textbf{Implementation (Chisel):} Bitonic Sort algorithm (Listing~\ref{code:sort_scala})

\vspace{6pt}
\begin{lstlisting}[language=C++, numbers=none]
#define N 8
extern "C" void Sort(unsigned char input[N], unsigned char output[N]) {
    // Sorts the 'input' array, stores the result in 'output'
    ...
}
\end{lstlisting}
\vspace{-10pt}
\captionof{listing}{Sort.cpp (Bubble Sort)}
\label{code:sort_cpp}
\vspace{6pt}

\begin{lstlisting}[language=scala, numbers=none]
class Sort(width: Int = 8) extends RawModule {
    val input = IO(Input(Vec(8, UInt(width.W))))
    val output = IO(Output(Vec(8, UInt(width.W))))
    // Sorts the 'io.input', drives the result to 'io.output'
    ...
}
\end{lstlisting}
\vspace{-10pt}
\captionof{listing}{Sort.scala (Bitonic Sort)}
\label{code:sort_scala}

\vspace{2.5pt}
\noindent\textit{5.1.2  Verification Process and Results.} 
We evaluate \textsc{EquivFusion} by conducting verification across two distinct scenarios.

\textbf{Experiment I: Consistent Sorting Order Verification.}
In the baseline scenario, both the C++ specification and the Chisel implementation are configured to sort the input array in ascending order. The backend solver returns \textit{UNSAT}, proving the Chisel implementation is functionally equivalent to the C++ specification, despite the fundamental algorithmic disparity (Bubble Sort \textit{vs.} Bitonic Sort).

\textbf{Experiment II: Discrepancy Detection (Bug Injection).}
Subsequently, to validate the tool's error detection capability, we modify the C++ specification to sort in descending order while leaving the Chisel implementation unchanged (ascending). The solver correctly returns \textit{SAT}, successfully flagging the semantic mismatch and identifying the designs as non-equivalent.

\subsection{Example 2: PyTorch \textit{v.s.} Gate-Level Netlist} \label{sec:pytorch_netlist}

\vspace{2pt}
\noindent\textit{5.2.1  Experimental Setup.} 
This case study targets the verification of a fundamental deep learning primitive: a dot product operator applied to two 2-dimensional, 8-bit integer vectors.

\textbf{Specification(Pytorch): } We define a high-level \texttt{torch.nn.Module} that encapsulates the standard \texttt{torch.dot} operation (Listing \ref{code:dot2_py}). 
To bridge the gap between Python-based deep learning frameworks and hardware verification, we leverage the \texttt{torch-mlir} compiler infrastructure. This pipeline automatically lowers the PyTorch module into the MLIR \textbf{Affine dialect} (Listing \ref{code:dot2_mlir}), preserving the loop structures and memory access patterns required for formal analysis.

\vspace{6pt}

\begin{lstlisting}[language=Python, numbers=none]
class DotModule(torch.nn.Module):
    def mm(self, a, b):
        return torch.dot(a, b)

    def forward(self, a, b):
        return self.mm(a, b)
\end{lstlisting}
\vspace{-10pt}
\captionof{listing}{dot.py}
\label{code:dot2_py}
\vspace{6pt}

\begin{lstlisting}[numbers=none]
module {
    func.func @dot(%arg0: memref<2xi8>, %arg1: memref<2xi8>) -> memref<i8> {
        ....
        affine.for %arg2 = 0 to 2 {
            %2 = affine.load %arg0[%arg2] : memref<2xi8>
            %3 = affine.load %arg1[%arg2] : memref<2xi8>
            %4 = arith.muli %2, %3 : i8
            affine.store %4, %alloc[%arg2] : memref<2xi8>
        }
        ...
        return %alloc_1 : memref<i8>
    }
}
\end{lstlisting}
\vspace{-10pt}
\captionof{listing}{dot.mlir}
\label{code:dot2_mlir}
\vspace{6pt}

\textbf{Implementation (Netlist): } The hardware implementation begins as a hand-written Register-Transfer Level (RTL) Verilog design (Listing \ref{code:dot2_verilog}). 
To simulate a realistic backend flow, we employ the open-source synthesis suite Yosys to compile RTL into a flattened gate-level netlist. The synthesis targets \textit{cmos\_cells.lib}, a representative standard cell library containing fundamental combinational logic gates, resulting in the structural netlist shown in Listing \ref{code:dot2_netlist_verilg}.

\vspace{6pt}
\begin{lstlisting}[language=SystemVerilogCustom, numbers=none]
module dot2_comb #(
    parameter N       = 2,   // Vector length
    parameter W       = 8,   // Bit-width of element
    parameter ACC_W   = 32,  // Bit-width of accumulator
    parameter O_W = 8        // Bit-width of Output
)(
    input  logic signed [N-1:0] [W-1:0] arg_0,
    input  logic signed [N-1:0] [W-1:0] arg_1,
    output logic signed [0:0][W-1:0] out_0
);
    logic signed [N-1:0] [(2*W)-1:0] prod;
    genvar i;
    generate
        for (i = 0; i < N; i = i + 1) begin : GEN_PROD
            assign prod[i] = $signed(arg_0[i]) * $signed(arg_1[i]);
        end
    endgenerate
    logic signed [ACC_W-1:0] sum_reg;
    ...   // Accumulate all products into sum_reg.
    assign out_0 = sum_reg[O_W-1:0];
endmodule
\end{lstlisting}
\vspace{-10pt}
\captionof{listing}{dot.v}
\label{code:dot2_verilog}

\vspace{6pt}
\begin{lstlisting}[language=SystemVerilogCustom, numbers=none]
module dot2_comb(arg_0, arg_1, out_0);
    wire _0000_;
    wire _0001_;
    ...
    NOT _0852_ (.A(arg_0[8]), .Y(_0307_));
    ...
endmodule
\end{lstlisting}
\vspace{-10pt}
\captionof{listing}{netlist.v}
\label{code:dot2_netlist_verilg}

\vspace{2.5pt}
\noindent\textit{5.2.2  Verification Process and Results.} 
We apply \textsc{EquivFusion} to verify the equivalence between the \textit{Affine MLIR (Spec.)} and the synthesized \textit{Netlist (Impl.)} across two distinct precision configurations:

\textbf{Experiment I: 8-bit Output Verification.}
In the setup, both the PyTorch specification and the Verilog implementation are configured with 8-bit inputs and outputs. The solver returns \textit{UNSAT}, proving the gate-level netlist is functionally equivalent to the high-level PyTorch specification within the constrained 8-bit domain.

\textbf{Experiment II: 32-bit Output Verification (Precision Mismatch).}
We extend both designs to 32-bit outputs, upon which the solver returns \textit{SAT}, indicating non-equivalence. \textsc{EquivFusion} pinpoints the root cause to the Verilog implementation's sign-extension of intermediate signals: the hardware design sign-extends intermediate values during the calculation, creating a subtle semantic divergence from the torch specification.

\section{Conclusion}
\label{sec:conclusion}
This paper presents \textsc{EquivFusion}, a unified equivalence checking framework that enables formal reasoning across heterogeneous design modalities, from high-level algorithmic specifications to hardware implementations.
Built on MLIR and CIRCT, \textsc{EquivFusion} provides an automated, solver-agnostic pipeline that systematically lowers diverse inputs into a common logic representation, eliminating the need for manually maintained reference models.
By supporting equivalence checking early in the design flow, the framework facilitates shift-left verification for datapath-intensive designs.
As an open-source tool, \textsc{EquivFusion} establishes the first step towards cross-layer formal verification. We plan to integrate  it into the CIRCT ecosystem to support standardized equivalence checking within MLIR-based hardware compilation flows.

\bibliographystyle{ACM-Reference-Format}
\bibliography{main}

\newpage
\appendix
\section{Walk through EquivFusion}

A recorded walkthrough is available at \url{https://youtu.be/AdEeZHU54qA}.

\subsection{Installation}
\textsc{EquivFusion} is compatible with Linux and macOS environments.
The following steps(Listing \ref{app:build_and_installl_equivfusion_cmds}) outline the procedure to clone the repository, configure the build system using CMake and Ninja, and install the required solvers (including AIGER, Bitwuzla and Kissat):

\vspace{6pt}
\begin{lstlisting}[language=bash, numbers=none]
# Clone the repository
git clone git@github.com:FORMiND-Lab/EquivFusion.git
cd EquivFusion

# Configure and build the project
mkdir build && cd build
cmake -G Ninja ..

# Build
ninja

# Install solvers
ninja install_solvers

# Add 'EquivFusion/build/bin' to your PATH environment variable
export PATH="$PWD/bin/:$PATH"
\end{lstlisting}
\vspace{-10pt}
\captionof{listing}{Commands to Build and Install EquivFusion}
\label{app:build_and_installl_equivfusion_cmds}
\vspace{6pt}

Building \textsc{EquivFusion} requires git, ninja, python3, cmake, a C++ toolchain, and the readline library, along with the z3 solver in the system PATH.

\subsection{Docker}    \label{app:docker}
To facilitate reproducibility and ease of deployment, we provide a Docker-based environment. The Docker image can be built and run using the following commands(Listing \ref{app:build_and_run_docker}):

\vspace{6pt}
\begin{lstlisting}[language=bash, numbers=none]
$ docker build -t equivfusion .
$ docker run -it equivfusion
\end{lstlisting}
\vspace{-10pt}
\captionof{listing}{Commands to Build and Run Docker}
\label{app:build_and_run_docker}

\subsection{Run Examples} \label{app:run_example}

\subsubsection{C++ \textit{v.s.} Chisel}
This appendix provides the complete source code and execution workflow for the C++ Specification vs Chisel Implementation case study presented in Section \ref{sec:cpp_chisel}.

\textbf{Specification(C++):} The high-level specification is written in C++ (Listing \ref{app:app_sort_cpp}), implementing the Bubble Sort algorithm:

\vspace{6pt}
\begin{lstlisting}[language=C++, numbers=none]
// Sort.cpp
#include <cstdint>

#define N 8

extern "C" void Sort(unsigned char input[N], unsigned char output[N]) {
    unsigned char temp[N];

    for (unsigned int i = 0; i < N; i++) {
        temp[i] = input[i];
    } 

    for (unsigned int i = N - 1; i > 0; i--) {
        unsigned char high = temp[0];
        unsigned char low = 0;

        for (unsigned int j = 1; j < N; j++) {
            if (j <= i) {
                if (temp[j] > high) {
                    low = high;
                    high = temp[j];
                } else {
                    low = temp[j];
                }
            } else {
	        low = temp[j - 1];
	    }

            temp[j - 1] = low;
        }

        temp[i] = high;
    }

    for (unsigned int i = 0; i < N; i++) {
        output[i] = temp[i];
    }
}
\end{lstlisting}
\vspace{-10pt}
\captionof{listing}{Sort.cpp}
\label{app:app_sort_cpp}
\vspace{6pt}

This C++ source is lowered into the MLIR Affine dialect(Listing \ref{app:app_sort_mlir}) using Polygeist:

\vspace{6pt}
\begin{lstlisting}[numbers=none]
// Sort.mlir
module {
  func.func @Sort(%arg0: memref<8xi8> {polygeist.param_name = "input"}, %arg1: memref<8xi8> {polygeist.param_name = "output"}) attributes {llvm.linkage = #llvm.linkage<external>} {
    %c8 = arith.constant 8 : index
    %alloca = memref.alloca() : memref<8xi8>
    affine.for %arg2 = 0 to 8 {
      %0 = affine.load %arg0[%arg2] : memref<8xi8>
      affine.store %0, %alloca[%arg2] : memref<8xi8>
    }
    affine.for %arg2 = 1 to 8 {
      %0 = arith.subi %c8, %arg2 : index
      %1 = arith.index_cast %0 : index to i32
      %2 = affine.load %alloca[0] : memref<8xi8>
      %3 = affine.for %arg3 = 1 to 8 iter_args(%arg4 = %2) -> (i8) {
        %4 = arith.index_cast %arg3 : index to i32
        %5 = arith.cmpi ule, %4, %1 : i32
        %6:2 = scf.if %5 -> (i8, i8) {
          %7 = affine.load %alloca[%arg3] : memref<8xi8>
          %8 = arith.extui %7 : i8 to i32
          %9 = arith.extui %arg4 : i8 to i32
          %10 = arith.cmpi sgt, %8, %9 : i32
          %11 = arith.select %10, %arg4, %7 : i8
          %12 = arith.select %10, %7, %arg4 : i8
          scf.yield %11, %12 : i8, i8
        } else {
          %7 = affine.load %alloca[%arg3 - 1] : memref<8xi8>
          scf.yield %7, %arg4 : i8, i8
        }
        affine.store %6#0, %alloca[%arg3 - 1] : memref<8xi8>
        affine.yield %6#1 : i8
      }
      affine.store %3, %alloca[-%arg2 + 8] : memref<8xi8>
    }
    affine.for %arg2 = 0 to 8 {
      %0 = affine.load %alloca[%arg2] : memref<8xi8>
      affine.store %0, %arg1[%arg2] : memref<8xi8>
    }
    return
  }
}
\end{lstlisting}
\vspace{-10pt}
\captionof{listing}{Sort.mlir}
\label{app:app_sort_mlir}
\vspace{6pt}

\textbf{Implementation(Chisel):} The hardware implementation is described in Chisel (Listing \ref{app:app_sort_scala}), implementing a Bitonic Sorter:

\vspace{6pt}
\begin{lstlisting}[language=scala, numbers=none]
// Sort.scala
import chisel3._
import circt.stage.ChiselStage

class CompareAndSwap(width: Int) extends RawModule {
  val io = IO(new Bundle {
    val a = Input(UInt(width.W))
    val b = Input(UInt(width.W))
    val min = Output(UInt(width.W))
    val max = Output(UInt(width.W))
  })

  when(io.a <= io.b) {
    io.min := io.a
    io.max := io.b
  }.otherwise {
    io.min := io.b
    io.max := io.a
  }
}

class Sort(width: Int = 8) extends RawModule {
  val input = IO(Input(Vec(8, UInt(width.W))))
  val output = IO(Output(Vec(8, UInt(width.W))))

  def CAS(a: UInt, b: UInt): (UInt, UInt) = {
    val m = Module(new CompareAndSwap(width))
    m.io.a := a
    m.io.b := b
    (m.io.min, m.io.max)
  }

  // === Stage 1 ===
  val (s1_0, s1_1) = CAS(input(0), input(1))
  val (s1_3, s1_2) = CAS(input(2), input(3)) // Swap order for bitonic merge
  val (s1_4, s1_5) = CAS(input(4), input(5))
  val (s1_7, s1_6) = CAS(input(6), input(7)) // Swap order

  // === Stage 2 ===
  val (s2_0_t, s2_2_t) = CAS(s1_0, s1_2)
  val (s2_1_t, s2_3_t) = CAS(s1_1, s1_3)
  val (s2_0, s2_1) = CAS(s2_0_t, s2_1_t)
  val (s2_2, s2_3) = CAS(s2_2_t, s2_3_t)

  val (s2_6_t, s2_4_t) = CAS(s1_4, s1_6) // Descending merge
  val (s2_7_t, s2_5_t) = CAS(s1_5, s1_7)
  val (s2_5, s2_4) = CAS(s2_4_t, s2_5_t)
  val (s2_7, s2_6) = CAS(s2_6_t, s2_7_t)

  // === Stage 3 ===
  val (s3_0_t1, s3_4_t1) = CAS(s2_0, s2_4)
  val (s3_1_t1, s3_5_t1) = CAS(s2_1, s2_5)
  val (s3_2_t1, s3_6_t1) = CAS(s2_2, s2_6)
  val (s3_3_t1, s3_7_t1) = CAS(s2_3, s2_7)

  val (s3_0_t2, s3_2_t2) = CAS(s3_0_t1, s3_2_t1)
  val (s3_1_t2, s3_3_t2) = CAS(s3_1_t1, s3_3_t1)
  val (s3_4_t2, s3_6_t2) = CAS(s3_4_t1, s3_6_t1)
  val (s3_5_t2, s3_7_t2) = CAS(s3_5_t1, s3_7_t1)

  val (o0, o1) = CAS(s3_0_t2, s3_1_t2)
  val (o2, o3) = CAS(s3_2_t2, s3_3_t2)
  val (o4, o5) = CAS(s3_4_t2, s3_5_t2)
  val (o6, o7) = CAS(s3_6_t2, s3_7_t2)

  output(0) := o0
  output(1) := o1
  output(2) := o2
  output(3) := o3
  output(4) := o4
  output(5) := o5
  output(6) := o6
  output(7) := o7
}

object Sort extends App {
  (new ChiselStage).execute(args, Seq(chisel3.stage.ChiselGeneratorAnnotation(() => new Sort(8))))
}
\end{lstlisting}
\vspace{-10pt}
\captionof{listing}{Sort.scala}
\label{app:app_sort_scala}
\vspace{6pt}

This design is compiled into the FIRRTL intermediate representation (Listing \ref{app:app_sort_fir}) using sbt:

\vspace{6pt}
\begin{lstlisting}[numbers=none]
FIRRTL version 3.3.0
circuit Sort :%[[
  {
    "class":"firrtl.transforms.DedupGroupAnnotation",
    "target":"~Sort|CompareAndSwap",
    "group":"CompareAndSwap"
  },

  ......
  
  {
    "class":"firrtl.transforms.DedupGroupAnnotation",
    "target":"~Sort|Sort",
    "group":"Sort"
  }
]]
  module CompareAndSwap : 
    output io : { flip a : UInt<8>, flip b : UInt<8>, min : UInt<8>, max : UInt<8>}

    node _T = leq(io.a, io.b) 
    when _T : 
      connect io.min, io.a 
      connect io.max, io.b 
    else :
      connect io.min, io.b 
      connect io.max, io.a 

......

  module Sort : 
    input input : UInt<8>[8] 
    output output : UInt<8>[8] 

    inst m of CompareAndSwap 
    connect m.io.a, input[0]
    connect m.io.b, input[1] 

  ......

    connect output[4], m_22.io.min 
    connect output[5], m_22.io.max 
    connect output[6], m_23.io.min 
    connect output[7], m_23.io.max 
\end{lstlisting}
\vspace{-10pt}
\captionof{listing}{Sort.fir}
\label{app:app_sort_fir}
\vspace{6pt}

\textbf{Execution and Results:} To perform the equivalence check, the following commands(Listing \ref{app:app_command_c_chisel}) are executed in \textit{equiv\_fusion}:

\vspace{6pt}
\begin{lstlisting}[language=bash, numbers=none]
read_c -spec -top Sort Sort.mlir
read_fir -impl -top Sort Sort.fir --scalarize-public-modules false --scalarize-public-modules false --preserve-aggregate all
equiv_miter -specModule Sort -implModule Sort -mitermode aiger -o miter.aiger
solver_runner --solver kissat --inputfile miter.aiger
\end{lstlisting}
\vspace{-10pt}
\captionof{listing}{Commands to Run C++ \textit{v.s.} Chisel Equivalence Checking}
\label{app:app_command_c_chisel}
\vspace{6pt}

The solver returns \textit{UNSAT}, proving that the Chisel implementation is functionally equivalent to the C++ specification despite algorithmic differences.

\subsubsection{Pytorch \textit{v.s.} Netlist}
This appendix details the complete source artifacts and verification workflow for the PyTorch Specification vs Gate-Level Netlist case study presented in Section\ref{sec:pytorch_netlist}.

\textbf{Specification(Pytorch):} The high-level specification is defined as a PyTorch module (Listing \ref{app:dot_py}) performing a dot product operation:

\vspace{6pt}
\begin{lstlisting}[language=Python, numbers=none]
# dot.py
from typing import List

import torch
import torch.nn as nn

from torch_mlir import fx

import os
import sys

class DotModule(torch.nn.Module):
  def mm(self, a, b):
    return torch.dot(a, b)
  
  def forward(self, a, b):
    return self.mm(a, b)
  
model = DotModule()

x = torch.randint(low=-128, high=128, size=(2,), dtype=torch.int8)
y = torch.randint(low=-128, high=128, size=(2,), dtype=torch.int8)

m = fx.export_and_import(model, x, y, output_type="linalg-on-tensors", func_name="dot")

current_script_path = os.path.abspath(__file__)
current_script_dir = os.path.dirname(current_script_path)

with open(os.path.join(current_script_dir, "dot.mlir"), "w") as f:
    f.write(str(m))
\end{lstlisting}
\vspace{-10pt}
\captionof{listing}{dot.py}
\label{app:dot_py}
\vspace{6pt}

This PyTorch model is lowered into the MLIR Affine dialect using the torch-mlir infrastructure combined with upstream MLIR tools. The resulting MLIR code (Listing \ref{app:app_dot_mlir}) serves as the input specification for \textsc{EquivFusion}:

\vspace{6pt}
\begin{lstlisting}[numbers=none]
// dot.mlir
module {
  func.func @dot(%arg0: memref<2xi8>, %arg1: memref<2xi8>) -> memref<i8> {
    %c0_i64 = arith.constant 0 : i64
    %alloc = memref.alloc() {alignment = 64 : i64} : memref<2xi8>
    affine.for %arg2 = 0 to 2 {
      %2 = affine.load %arg0[%arg2] : memref<2xi8>
      %3 = affine.load %arg1[%arg2] : memref<2xi8>
      %4 = arith.muli %2, %3 : i8
      affine.store %4, %alloc[%arg2] : memref<2xi8>
    }
    %alloc_0 = memref.alloc() {alignment = 64 : i64} : memref<i64>
    affine.store %c0_i64, %alloc_0[] : memref<i64>
    affine.for %arg2 = 0 to 2 {
      %2 = affine.load %alloc[%arg2] : memref<2xi8>
      %3 = affine.load %alloc_0[] : memref<i64>
      %4 = arith.extsi %2 : i8 to i64
      %5 = arith.addi %4, %3 : i64
      affine.store %5, %alloc_0[] : memref<i64>
    }
    %alloc_1 = memref.alloc() {alignment = 64 : i64} : memref<i8>
    %0 = affine.load %alloc_0[] : memref<i64>
    %1 = arith.trunci %0 : i64 to i8
    affine.store %1, %alloc_1[] : memref<i8>
    return %alloc_1 : memref<i8>
  }
}
\end{lstlisting}
\vspace{-10pt}
\captionof{listing}{dot.mlir}
\label{app:app_dot_mlir}
\vspace{6pt}

\textbf{Implementation(Netlist):} The hardware implementation originates from a SystemVerilog design (Listing \ref{app:app_dot_v}):

\vspace{6pt}
\begin{lstlisting}[language=SystemVerilogCustom, numbers=none]
// dot.v
module dot2_comb #(
    parameter N       = 2,   // Vector length
    parameter W       = 8,   // Bit-width of each element (signed)
    parameter ACC_W   = 32,    // Bit-width of accumulator
    parameter O_W = 8          // Bit-width of Output
)(
    // Two signed input vectors, each containing N elements of W bits
    input  logic signed [N-1:0] [W-1:0] arg_0,
    input  logic signed [N-1:0] [W-1:0] arg_1,

    // Signed output: dot product result (sum of element-wise products)
    output logic signed [0:0][W-1:0] out_0
);

    //===============================================
    // 1. Element-wise multiplications
    // Each pair a[i], b[i] is multiplied to produce a 2*W-bit product.
    //===============================================
    logic signed [N-1:0] [(2*W)-1:0] prod;

    genvar i;
    generate
        for (i = 0; i < N; i = i + 1) begin : GEN_PROD
            assign prod[i] = $signed(arg_0[i]) * $signed(arg_1[i]);
        end
    endgenerate

    //===============================================
    // 2. Summation of all products
    // This always block performs a combinational reduction (sum)
    // across all products. The result is stored in sum_reg.
    //===============================================
    integer k;
    logic signed [ACC_W-1:0] sum_reg;

    always @(*) begin
        sum_reg = '0; // Initialize accumulator to zero
        for (k = 0; k < N; k = k + 1) begin
            // Sign-extend each 2*W-bit product to ACC_W bits
            // before adding, to prevent overflow and preserve sign.
            sum_reg = sum_reg +
                {{(ACC_W-(2*W)){prod[k][(2*W)-1]}}, prod[k]};
        end
    end

    //===============================================
    // 3. Final output assignment
    // The dot product is simply the total accumulated sum.
    //===============================================
    assign out_0 = sum_reg[O_W-1:0];

endmodule
\end{lstlisting}
\vspace{-10pt}
\captionof{listing}{dot.v}
\label{app:app_dot_v}
\vspace{6pt}

To synthesize this design into a gate-level netlist, we utilize Yosys with a standard cell library \textit{cmos\_cells.lib}(Listing \ref{app:app_cmos_cells_lib}) provided in the Yosys examples. The corresponding Verilog simulation model for the cells is \textit{cmos\_cells.v}(Listing \ref{app:app_cmos_cells_v}):

\vspace{6pt}
\begin{lstlisting}[numbers=none]
// cmos_cells.lib
library(demo) {
  cell(BUF) {
    area: 6;
    pin(A) { direction: input; }
    pin(Y) { direction: output;
              function: "A"; }
  }
  cell(NOT) {
    area: 3;
    pin(A) { direction: input; }
    pin(Y) { direction: output;
              function: "A'"; }
  }
  cell(NAND) {
    area: 4;
    pin(A) { direction: input; }
    pin(B) { direction: input; }
    pin(Y) { direction: output;
             function: "(A*B)'"; }
  }
  cell(NOR) {
    area: 4;
    pin(A) { direction: input; }
    pin(B) { direction: input; }
    pin(Y) { direction: output;
             function: "(A+B)'"; }
  }
  cell(DFF) {
    area: 18;
    ff(IQ, IQN) { clocked_on: C;
                  next_state: D; }
    pin(C) { direction: input;
                 clock: true; }
    pin(D) { direction: input; }
    pin(Q) { direction: output;
              function: "IQ"; }
  }
  cell(DFFSR) {
    area: 18;
    ff("IQ", "IQN") { clocked_on: C;
                  next_state: D;
                      preset: S;
                       clear: R; }
    pin(C) { direction: input;
                 clock: true; }
    pin(D) { direction: input; }
    pin(Q) { direction: output;
              function: "IQ"; }
    pin(S) { direction: input; }
    pin(R) { direction: input; }
    ; // empty statement
  }
}
\end{lstlisting}
\vspace{-10pt}
\captionof{listing}{cmos\_cells.lib}
\label{app:app_cmos_cells_lib}
\vspace{6pt}

\vspace{6pt}
\begin{lstlisting}[language=Verilog, numbers=none]
// cmos_cells.v
module BUF(A, Y);
input A;
output Y;
assign Y = A;
endmodule

module NOT(A, Y);
input A;
output Y;
assign Y = ~A;
endmodule

module NAND(A, B, Y);
input A, B;
output Y;
assign Y = ~(A & B);
endmodule

module NOR(A, B, Y);
input A, B;
output Y;
assign Y = ~(A | B);
endmodule

module DFF(C, D, Q);
input C, D;
output reg Q;
always @(posedge C)
	Q <= D;
endmodule

module DFFSR(C, D, Q, S, R);
input C, D, S, R;
output reg Q;
always @(posedge C, posedge S, posedge R)
	if (S)
		Q <= 1'b1;
	else if (R)
		Q <= 1'b0;
	else
		Q <= D;
endmodule
\end{lstlisting}
\vspace{-10pt}
\captionof{listing}{cmos\_cells.v}
\label{app:app_cmos_cells_v}
\vspace{6pt}

The synthesis is performed by executing the following Yosys passes(Listing \ref{app:app_yosys_netlist_cmds}):

\vspace{6pt}
\begin{lstlisting}[language=bash, numbers=none]
read_verilog -sv dot.v
hierarchy -check -top top

proc; opt; fsm; opt; memory; opt

techmap; opt

dfflibmap -liberty cmos_cells.lib

abc -liberty cmos_cells.lib

opt_clean

write_verilog netlist.v
\end{lstlisting}
\vspace{-10pt}
\captionof{listing}{Yosys Passes to Synthesized Netlist}
\label{app:app_yosys_netlist_cmds}
\vspace{6pt}

This process generates the gate-level netlist (Listing \ref{app:app_netlist_v}). A truncated snippet of the netlist is shown below:

\vspace{6pt}
\begin{lstlisting}[language=Verilog, numbers=none]
// netlist.v
/* Generated by Yosys 0.53+98 (git sha1 50b63c648, g++ 13.3.0-6ubuntu2~24.04 -Og -fPIC) */

(* dynports =  1  *)
(* top =  1  *)
(* src = "../../verilog/dot2_8/dot2.v:1.1-52.10" *)
module dot2_comb(arg_0, arg_1, out_0);
  wire _0000_;
  wire _0001_;
  wire _0002_;
  wire _0003_;
  wire _0004_;
  wire _0005_;
  wire _0006_;
  wire _0007_;
  wire _0008_;
  wire _0009_;
  wire _0010_;

  ......

  wire [31:0] sum_reg;
  NOT _0852_ (
    .A(arg_0[8]),
    .Y(_0307_)
  );
  NOT _0853_ (
    .A(arg_1[15]),
    .Y(_0318_)
  );
  NAND _0854_ (
    .A(arg_0[0]),
    .B(arg_1[0]),
    .Y(_0329_)
  );
  NAND _0855_ (
    .A(arg_0[8]),
    .B(arg_1[8]),
    .Y(_0340_)
  );

  ......

    NOT _1710_ (
    .A(_0840_),
    .Y(_0841_)
  );
  NOR _1711_ (
    .A(_0351_),
    .B(_0841_),
    .Y(out_0[0])
  );
  assign k = 32'd2;
  assign sum_reg[7:0] = out_0;
endmodule
\end{lstlisting}
\vspace{-10pt}
\captionof{listing}{netlist.v}
\label{app:app_netlist_v}
\vspace{6pt}

\textbf{Execution and Results:} To verify the equivalence between the generated MLIR specification and the synthesized netlist, the following commands(Listing \ref{app:app_command_pytorch_netlist}) are executed in \textit{equiv\_fusion}:

\vspace{6pt}
\begin{lstlisting}[language=bash, numbers=none]
read_c -spec -top dot dot.mlir
read_v -impl -top dot2_comb netlist.v cmos_cells.v
equiv_miter -specModule dot -implModule dot2_comb -mitermode btor2 -o miter.btor2
solver_runner --solver bitwuzla --inputfile miter.btor2
\end{lstlisting}
\vspace{-10pt}
\captionof{listing}{Commands to Run Pytorch \textit{v.s.} Netlist Equivalence Checking}
\label{app:app_command_pytorch_netlist}
\vspace{6pt}

The result \textit{UNSAT} confirms functional equivalence under the 8-bit configuration.

\subsection{Debugging}
To ensure robustness and facilitate development, \textsc{EquivFusion} provides a rigorous testing infrastructure and fine-grained debugging utilities:

\textbf{Continuous Integration (CI): }A GitHub Actions pipeline enforces code stability by automatically validating build success and executing a comprehensive suite of integration tests on every pull request, preventing functional regression.

\textbf{equivfusion-hls: }This utility isolates the HLS flow execution, enabling developers to inspect the intermediate MLIR IR generated after each transformation pass for deep analysis of the lowering process.

\textbf{equivfusion-opt: }Modeled after mlir-opt, this tool allows for the isolated invocation of individual EquivFusion passes, facilitating targeted debugging and the development of custom optimization logic.

\section{Runtime Analysis}
Table \ref{tab:execution_time_table} details the execution time for each command in the verification workflows of the examples presented in Appendix \ref{app:run_example}.

\begin{table}[htbp]
\centering
\caption{Execution Time(seconds)}  
\label{tab:execution_time_table} 
\begin{tabular}{@{}lcccc@{}}
\toprule
\textbf{Stage}              & \textbf{\makecell{bitwuzla\\smt}} & \textbf{\makecell{bitwuzla\\btor2}}   & \textbf{\makecell{kissat\\aiger}}   \\ \bottomrule
\rowcolor[HTML]{F2F3F5}
\multicolumn{4}{c}{\textbf{Execution Time (UNSAT): Sort.scala vs Sort.cpp}}                  \\ \bottomrule
\textbf{set\_port}          & 0.000006      & 0.000006      & 0.000006                      \\
\textbf{read\_c}            & 0.017226      & 0.013448      & 0.013090                      \\
\textbf{read\_firrtl}       & 0.060061      & 0.073934      & 0.067827                      \\
\textbf{equiv\_miter}       & 0.010399      & 0.008282      & 0.031100                      \\
\textbf{solver\_runner}     & 47.54907      & 55.82388      & 45.67969                      \\ \bottomrule
\rowcolor[HTML]{F2F3F5} 
\multicolumn{4}{c}{\textbf{Execution Time (SAT): Sort.scala vs Sort.cpp}}                    \\ \bottomrule
\textbf{set\_port}          & 0.000005      & 0.000006      & 0.000005                      \\
\textbf{read\_c}            & 0.013120      & 0.012417      & 0.012875                      \\
\textbf{read\_firrtl}       & 0.055784      & 0.079103      & 0.052333                      \\
\textbf{equiv\_miter}       & 0.006621      & 0.007322      & 0.027125                      \\
\textbf{solver\_runner}     & 0.007440      & 0.013190      & 0.018456                      \\ \bottomrule
\rowcolor[HTML]{F2F3F5}
\multicolumn{4}{c}{\textbf{Execution Time (UNSAT): Dot.py vs netlist.v}}                         \\ \bottomrule
\textbf{read\_c}            & 0.003391      & 0.003289      & 0.003449                      \\
\textbf{read\_v}            & 0.049022      & 0.049296      & 0.052890                      \\
\textbf{equiv\_miter}       & 0.023631      & 0.013858      & 0.022719                      \\
\textbf{solver\_runner}     & 2.776137      & 3.013583      & 1.800442                      \\ \bottomrule
\rowcolor[HTML]{F2F3F5} 
\multicolumn{4}{c}{\textbf{Execution Time (SAT): Dot.py vs netlist.v}}                           \\ \bottomrule
\textbf{read\_c}            & 0.003769      & 0.003429      & 0.005344                      \\
\textbf{read\_v}            & 0.251780      & 0.278213      & 0.262784                      \\
\textbf{equiv\_miter}       & 0.058312      & 0.037827      & 0.066539                      \\
\textbf{solver\_runner}     & 0.039862      & 0.041944      & 0.055321                      \\ \bottomrule
\end{tabular}
\end{table}

\section{Synthesizable Subset and Constraints} \label{app:syntax_constraints}
For high-level languages like C/C++ and PyTorch, \textsc{EquivFusion} targets a specific synthesizable subset conducive to High-Level Synthesis (HLS), with the following specific constraints:

\textbf{Data Types:} Support is limited to integer arithmetic and fixed-width bit-vectors. Floating-point support is planned via future emulation library integration.

\textbf{Control Flow (Branching):} Standard conditionals (e.g., if-else) are supported if statically resolvable or mappable to multiplexers.

\textbf{Control Flow (Loops):} Loop support is strictly limited to static affine \textit{for} loops with compile-time determinable bounds and constant strides. Unstructured control flow (e.g., \textit{break, continue}) is prohibited to ensure deterministic unrolling.

\textbf{Memory \& Pointers:} Memory operations use statically allocated arrays. Pointers require fixed sizes and explicit indexing, and pointer reassignment within conditional scopes is prohibited to ensure Static Single Assignment (SSA) consistency.

\section{More Cases}

\subsection{Dot64 (C++ v.s. Chisel)}
This case demonstrates the end-to-end \textsc{EquivFusion} workflow for verifying equivalence between a high-level algorithmic C++ model and a Chisel design, using the 64-element dot product (dot64) computation as a canonical benchmark.

\textbf{Specification (Chisel): } FIRRTL, derived from the Chisel design(Listing \ref{app:app_dot_64_scala}) via Chisel elaboration.

\vspace{6pt}
\begin{lstlisting}[language=scala, numbers=none]
class Dot64 extends RawModule {
    val arg_0 = IO(Input(Vec(64, SInt(16.W))))
    val arg_1 = IO(Input(Vec(64, SInt(16.W))))
    val out_0 = IO(Output(SInt(64.W)))
    var sum = 0.S(64.W)
    for (i <- 0 until 64) {
        val product = arg_0(i) * arg_1(i)
        sum = sum + product
    }
    out_0 := sum
}
\end{lstlisting}
\vspace{-10pt}
\captionof{listing}{Dot64.scala}
\label{app:app_dot_64_scala}
\vspace{6pt}

\textbf{Implementation (C++): } MLIR, derived from the high-level C++ algorithm via Polygeist.

\texttt{Dot64.cpp}(Listing \ref{app:app_dot_64_cpp}): Fully consistent with Chisel specification.

\vspace{6pt}
\begin{lstlisting}[language=C++, numbers=none]
extern "C" int64_t Dot64(const int16_t (&arg_0)[64], const int16_t (&arg_1)[64]) {
    int64_t sum = 0;
    for (int i = 0; i < 64; ++i) {
        int32_t product = arg_0[i] * arg_1[i];
        sum += product;
    }
    return sum;
}
\end{lstlisting}
\vspace{-10pt}
\captionof{listing}{Dot64.cpp}
\label{app:app_dot_64_cpp}
\vspace{6pt}

\texttt{Dot64\_truncation.cpp}(Listing \ref{app:app_dot_64_truncation_cpp}): Contains a bug: product truncation to 16 bits. 

\vspace{6pt}
\begin{lstlisting}[language=C++, numbers=none]
        int16_t product = arg_0[i] * arg_1[i];
\end{lstlisting}
\vspace{-10pt}
\captionof{listing}{Dot64\_truncation.cpp}
\label{app:app_dot_64_truncation_cpp}
\vspace{6pt}

\textbf{Verification Results: }
\begin{itemize}
    \item \textbf{Dot64.scala vs Dot64.cpp} \\ The solver returns UNSAT, proving functional equivalence.
	\item \textbf{Dot64.scala vs Dot64\_truncation.cpp} \\The solver returns SAT, indicating non-equivalence.
\end{itemize}

\section{Future Directions}
\subsection{Intelligent Solver Orchestration}
Building on the foundational \texttt{solver\_runner} module—which currently orchestrates Z3, Bitwuzla, and Kissat—future efforts will focus on evolving the orchestration engine from a static dispatcher into an adaptive, intelligent decision-making layer. We target three key enhancements:

\textbf{Expanded Verification Engine Integration:} 
We aim to incorporate a broader spectrum of domain-specific engines to cover diverse verification needs. This includes integrating \textbf{ABC} for optimized combinational equivalence checking and \textbf{CVC5} for handling more complex theories.

\textbf{Adaptive Solver Selection and Configuration:} 
Different solvers and configurations exhibit vastly different performance characteristics depending on the circuit structure (e.g., arithmetic-heavy vs. control-heavy). To optimize verification throughput, we plan to implement automated selection mechanisms:  \textbf{Rule-Based Heuristics:} Leveraging static analysis of the IR to extract circuit features (e.g., bit-width distribution, logic depth, operator types). Based on these signatures, the system will apply expert rules to dispatch tasks (e.g., routing purely boolean logic to AIGER-based SAT solvers like Kissat, while directing complex word-level arithmetic to Bitwuzla).
\textbf{AI-Driven Optimization:} We explore employing Machine Learning (ML) models, such as Graph Neural Networks (GNNs), to embed circuit netlists and predict the most efficient solver for a given instance. Furthermore, we intend to utilize Bayesian Optimization or Reinforcement Learning to automatically tune solver hyperparameters (e.g., restart strategies, decision heuristics) dynamically, minimizing solving time without manual intervention.

\textbf{Parallel Portfolio Execution:} 
Beyond sequential invocation, we will implement a parallel portfolio strategy. This involves launching multiple solvers (or the same solver with distinct random seeds/configurations) concurrently on multi-core systems. The orchestration layer will terminate all processes as soon as the fastest solver returns a result, thereby reducing the latency of the "straggler" effect in hard verification instances.

\subsection{Floating-Point Verification via Soft-Float Emulation}
While the current iteration of \textsc{EquivFusion} focuses on integer arithmetic, extending verification capabilities to floating-point domains is a key objective. We propose integrating a canonical floating-point emulation library (e.g., SoftFloat) to bridge this gap. The envisaged approach involves lowering high-level floating-point IR operations (e.g., \texttt{arith.addf}, \texttt{arith.mulf}) into function calls that invoke verified software implementations of IEEE-754 compliant operators. By replacing native floating-point instructions with bit-precise integer logic from the emulation library, \textsc{EquivFusion} will enable rigorous bit-level equivalence checking between algorithmic specifications and hardware FPUs. This strategy effectively bypasses the complexity of floating-point theories, allowing standard bit-vector solvers to reason about floating-point semantics.

\subsection{Supporting Hand-Written Assumes and Lemmas}
Although \textsc{EquivFusion} targets fully automated equivalence checking, the semantic gap between high-level algorithms and low-level RTL can sometimes exceed the capabilities of automatic inference—particularly when dealing with complex loop unrolling, retiming, or aggressive synthesis optimizations. To address these challenges, a promising future direction is to support \textbf{user-guided verification} through hand-written \textit{assumes} and \textit{lemmas}. \textbf{Assumes} define critical environmental constraints or input invariants (e.g., valid control signal ranges or loop bounds) under which equivalence is expected to hold, effectively pruning the solver's search space. \textit{Lemmas} capture intermediate semantic properties, such as algorithmic invariants or correspondence points between pipeline stages. These can serve as "checkpoints" to guide the solver in state alignment.

This proposed workflow draws inspiration from the methodology employed in Synopsys Hector~\cite{synopsys_vcformal_dpv}. 
Similar to Hector's approach of leveraging user insights to decompose monolithic proofs, \textsc{EquivFusion} aims to translate these user-provided annotations into solver constraints and proof sub-goals. 
By incorporating verified lemmas to guide equivalence partitioning and cross-level state alignment, we seek to combine the scalability of automated solving with the precision of manual guidance, without altering the core verification pipeline.

\end{document}